\documentclass[aps,amsfonts,nofootinbib,preprintnumbers,nobalancelastpage]{revtex4}

\usepackage[english]{babel}
\usepackage{amsmath,amssymb}
\usepackage[dvips]{graphicx}
\usepackage[sort&compress]{natbib}
\usepackage{bbm}
\usepackage{color}
\usepackage{ulem}

\newcommand{\ie} {{\it i.e.}}

\begin{document}

%------include the correct preprint number
%\preprint{}

\title{Constraining anomalous Higgs interactions}

\author{Tyler Corbett}
\email{corbett.ts@gmail.com}
\affiliation{%
  C.N.~Yang Institute for Theoretical Physics, SUNY at Stony Brook,
  Stony Brook, NY 11794-3840, USA}

\author{O.\ J.\ P.\ \'Eboli}
\email{eboli@fma.if.usp.br}
\affiliation{Instituto de F\'{\i}sica,
             Universidade de S\~ao Paulo, S\~ao Paulo -- SP, Brazil.}
\affiliation{Institut de Physique Th\'{e}orique, CEA-Saclay
Orme des Merisiers, 91191 Gif-sur-Yvette, France}

\author{J.\ Gonzalez--Fraile}
\email{fraile@ecm.ub.edu}
\affiliation{%
  Departament d'Estructura i Constituents de la Mat\`eria and
  ICC-UB, Universitat de Barcelona, 647 Diagonal, E-08028 Barcelona,
  Spain}

\author{M.\ C.\ Gonzalez--Garcia} \email{concha@insti.physics.sunysb.edu}
\affiliation{%
  C.N.~Yang Institute for Theoretical Physics, SUNY at Stony Brook,
  Stony Brook, NY 11794-3840, USA}
\affiliation{%
  Instituci\'o Catalana de Recerca i Estudis Avan\c{c}ats (ICREA),}
\affiliation {Departament d'Estructura i Constituents de la Mat\`eria, Universitat
  de Barcelona, 647 Diagonal, E-08028 Barcelona, Spain}

%---------------------------------------------------------------------
\begin{abstract}

  The recently announced Higgs discovery marks the dawn of the direct
  probing of the electroweak symmetry breaking sector. Sorting out the
  dynamics responsible for electroweak symmetry breaking now requires
  probing the Higgs interactions and searching for additional states
  connected to this sector. In this work we analyze the constraints on
  Higgs couplings to the standard model gauge bosons using the
  available data from Tevatron and LHC.  We work in a
  model--independent framework expressing the departure of the Higgs
  couplings to gauge bosons by dimension--six operators.  This allows
  for independent modifications of its couplings to gluons, photons
  and weak gauge bosons while still preserving the Standard Model (SM)
  gauge invariance. Our results indicate that best overall agreement
  with data is obtained if the cross section of Higgs production via
  gluon fusion is suppressed with respect to its SM value and the
  Higgs branching ratio into two photons is enhanced, while keeping
  the production and decays associated to couplings to weak gauge
  bosons close to their SM prediction.

\end{abstract}
%---------------------------------------------------------------------

%\pacs{} 

\maketitle

\renewcommand{\baselinestretch}{1.15}
%%%%%%%%%%%%%%%%%%%%%%%%%%%%%%%%%%%%%%%%%%%%%%%%%%%%%%%%%%%%%%%%%%%%%%
\section{Introduction}

The electroweak symmetry breaking (EWSB) mechanism has been elusive
for many decades. However the recently announced discovery of a 125
GeV Higgs boson~\cite{ Englert:1964et, Higgs:1964pj, Higgs:1964ia,
  Guralnik:1964eu, Higgs:1966ev, Kibble:1967sv} at the CERN Large
Hadron Collider (LHC) ~\cite{ atlas8summer, cms8summer} opens a new
era in particle physics.  The pressing questions now are related to
the properties of this new observed state, like its spin and
couplings, in order to extend our knowledge of the EWSB sector.  In
this work we employ the data used for the Higgs discovery to constrain
its couplings to gauge bosons.  \smallskip

Presently there are many possible EWSB scenarios ranging from the
Higgs being elementary and weakly interacting~\cite{Altarelli:2012dq},
as in the Standard Model, to it being composite and related to a new
strongly interacting sector~\cite{Dimopoulos:1979es,
  Weinberg:1979bn}.  In this last case the precision electroweak
measurements and flavor changing neutral currents lead to strong
constraints. However, recent theoretical advances have made possible
the construction of models in agreement with the experimental
bounds~\cite{Hill:2002ap}.  The distinction between the different
scenarios can be carried out by looking for further new states
associated with the EWSB mechanism and/or by careful studies of the
Higgs boson couplings. \smallskip

In this work we assume that the observed Higgs boson is part of a
$SU(2)_L$ doublet and that possible additional states are heavy enough
not to play a direct role in the low energy phenomenology.  This
assumption is realized in models where the Higgs boson is a
pseudo--Goldstone boson of a larger broken global
symmetry~\cite{Kaplan:1983fs, Kaplan:1983sm, Banks:1984gj,Dugan:1984hq,
  Georgi:1984ef, Agashe:2004rs, Giudice:2007fh%, Dimopoulos:1981xc
}. Under this assumption we consider the most general dimension--six
effective Lagrangian invariant under linear $SU(3)_c \otimes SU(2)_L
\otimes U(1)_Y$ transformations to describe the interactions of the
Higgs boson with the electroweak gauge bosons, as well as with the
gluons~\cite{Buchmuller:1985jz, Leung:1984ni}. For the sake of
simplicity we assume that the Higgs has the same interaction with
fermions as in the SM, nevertheless this hypothesis still has to be
tested further\footnote{The preliminary CMS~\cite{cms8summer} results
  indicate that the SM values for the Higgs couplings to fermions are
  within the 90--95\% CL allowed region.}. This scenario can be
falsified by the discovery of new states or by the non--observation of
its predictions to the triple electroweak--gauge--boson vertices.
\smallskip

The effective operators describing the Higgs anomalous interactions
modify both the Higgs production mechanisms and its decay patterns,
therefore we combine several channels to unravel the contribution of
the different operators.  In our analyses we use the most recent data
from the Tevatron~\cite{tevanew} and LHC at 7 TeV~\cite{atlas7new,
  Chatrchyan:2012tx} and at 8 TeV~\cite{atlas8summer, cms8summer,
  cmspashig12015, cmspashig12020, atlasconf12091}. Anomalous
interactions also enhance the Higgs decay into $Z\gamma$ as well as
its production in association with a photon. Nevertheless, the
available statistics is not enough to make these channels visible.
\smallskip

This article is organized as follows. In Section II we introduce the
dimension--six effective operators and the different scenarios studied
in this work. Details of our analyses are presented in Section III and
Section IV contains their results. Finally we discuss the main
conclusions in Section V. \smallskip

%%%%%%%%%%%%%%%%%%%%%%%%%%%%%%%%%%%%%%%%%%%%%%%%%%%%%%%%%%%%%%%%%%%%%%
\section{Higgs anomalous interactions}

In this work we assume that even if there is new physics associated
with the electroweak symmetry breaking sector, the Higgs boson
observed at LHC is still part of a $SU(2)_L$ doublet, the SM gauge
invariance holds and no additional light states, relevant to the Higgs
observables, are present in the spectrum.  Under these assumptions the
new effects can be parametrized in a model independent way by
extending the SM with the addition of higher dimension operators that
are invariant under linear $SU(3)_c \otimes SU(2)_L \otimes U(1)_Y$
transformations. \smallskip

In this framework the first corrections to the Higgs couplings to
gauge bosons are expressed as dimension--six operators that can be
written as
\begin{equation}
{\cal L}_{\text{eff}} = \sum_n \frac{f_n}{\Lambda^2} {\cal O}_n \;\; ,
\label{l:eff}
\end{equation}
where the operators ${\cal O}_n$ involve vector--boson and/or
Higgs--boson fields with couplings $f_n$ and where $\Lambda$ is a
characteristic scale.  Requiring the operators ${\cal O}_{n}$ to be
$P$ and $C$ even, there are only seven dimension--six operators that
modify the Higgs--boson couplings to electroweak vector bosons and one
to gluons~\cite{Buchmuller:1985jz,Leung:1984ni}:
\begin{equation}
\begin{array}{lll}
 {\cal O}_{GG} = \Phi^\dagger \Phi \; G_{\mu\nu}^a G^{a\mu\nu}  \;\;,
& {\cal O}_{WW} = \Phi^{\dagger} \hat{W}_{\mu \nu} 
 \hat{W}^{\mu \nu} \Phi  \;\; , 
& {\cal O}_{BB} = \Phi^{\dagger} \hat{B}_{\mu \nu} 
  \hat{B}^{\mu \nu} \Phi \;\; ,  
\\
& &
\\
 {\cal O}_{BW} =  \Phi^{\dagger} \hat{B}_{\mu \nu} 
 \hat{W}^{\mu \nu} \Phi \;\; ,
&
  {\cal O}_W  = (D_{\mu} \Phi)^{\dagger} 
  \hat{W}^{\mu \nu}  (D_{\nu} \Phi) \;\; ,

& {\cal O}_B  =  (D_{\mu} \Phi)^{\dagger} 
  \hat{B}^{\mu \nu}  (D_{\nu} \Phi)  \;\; ,
\\
& &
\\
 {\cal O}_{\Phi,1} = \left ( D_\mu \Phi \right)^\dagger \Phi^\dagger \Phi
\left ( D^\mu \Phi \right ) \;\; , 
&{\cal O}_{\Phi,2} = \frac{1}{2} 
\partial^\mu\left ( \Phi^\dagger \Phi \right)
\partial_\mu\left ( \Phi^\dagger \Phi \right)
 \;\; , 

\end{array}
\label{eff}  
\end{equation}
where $\Phi$ stands for the Higgs doublet, $D_\mu$ is the covariant
derivative, $\hat{B}_{\mu \nu} = i (g'/2) B_{\mu \nu}$ and
$\hat{W}_{\mu \nu} = i (g/2) \sigma^a W^a_{\mu \nu}$, with $B_{\mu
  \nu}$, $ W^a_{\mu \nu}$, and $G^a_{\mu\nu}$ being respectively the
$U(1)_Y$, $SU(2)_L$ and $SU(3)_c$ field strength tensors.  We denote
the $SU(2)_L$ ($U(1)_Y$) gauge coupling as $g$ ($g^\prime$) and the
Pauli matrices as $\sigma^a$.  \smallskip

The effective operators in Eq.\ (\ref{eff}) give rise to anomalous
$Hgg$, $H\gamma\gamma$, $HZ\gamma$, $HZZ$, and $HW^+W^-$ couplings,
which in the unitary gauge are given by
\begin{eqnarray}
{\cal L}_{\text{eff}}^{\text{HVV}} &=& 
g_{Hgg} \; H G^a_{\mu\nu} G^{a\mu\nu} +
g_{H \gamma \gamma} \; H A_{\mu \nu} A^{\mu \nu} + 
g^{(1)}_{H Z \gamma} \; A_{\mu \nu} Z^{\mu} \partial^{\nu} H + 
g^{(2)}_{H Z \gamma} \; H A_{\mu \nu} Z^{\mu \nu}
\nonumber \\
&+& g^{(1)}_{H Z Z}  \; Z_{\mu \nu} Z^{\mu} \partial^{\nu} H + 
g^{(2)}_{H Z Z}  \; H Z_{\mu \nu} Z^{\mu \nu} +
{g}^{(3)}_{H Z Z}  \; H Z_\mu Z^\mu  \\
\label{eff:nn}
&+& g^{(1)}_{H W W}  \; \left (W^+_{\mu \nu} W^{- \, \mu} \partial^{\nu} H 
+\text{h.c.} \right) +
g^{(2)}_{H W W}  \; H W^+_{\mu \nu} W^{- \, \mu \nu}
+g^{(3)}_{H W W}  \; H W^+_{\mu} W^{- \, \mu}
 \;\; ,\nonumber
\end{eqnarray}
where $V_{\mu \nu} = \partial_\mu V_\nu - \partial_\nu V_\mu$ with
$V=A$, $Z$ and $W$. The effective couplings $g_{Hgg}$, $g_{H \gamma
  \gamma}$, $g^{(1,2)}_{H Z \gamma}$, $g^{(1,2,3)}_{H W W}$ and
$g^{(1,2,3)}_{H Z Z}$ are related to the coefficients of the operators
appearing in (\ref{l:eff}) through,
\begin{eqnarray}
g_{Hgg} &=& \frac{f_{GG} v}{\Lambda^2}\equiv
-\frac{\alpha_s}{8 \pi} \frac{f_g v}{\Lambda^2} \;\;,
\nonumber \\
g_{H \gamma \gamma} &=& - \left( \frac{g M_W}{\Lambda^2} \right)
                       \frac{s^2 (f_{BB} + f_{WW} - f_{BW})}{2} \;\; , 
\nonumber \\
g^{(1)}_{H Z \gamma} &=& \left( \frac{g M_W}{\Lambda^2} \right) 
                     \frac{s (f_W - f_B) }{2 c} \;\; ,  
\nonumber \\
g^{(2)}_{H Z \gamma} &=& \left( \frac{g M_W}{\Lambda^2} \right) 
                      \frac{s [2 s^2 f_{BB} - 2 c^2 f_{WW} + 
                     (c^2-s^2)f_{BW} ]}{2 c}  \;\; , 
\nonumber \\ 
g^{(1)}_{H Z Z} &=& \left( \frac{g M_W}{\Lambda^2} \right) 
	              \frac{c^2 f_W + s^2 f_B}{2 c^2} \nonumber \;\; , \\
g^{(2)}_{H Z Z} &=& - \left( \frac{g M_W}{\Lambda^2} \right) 
  \frac{s^4 f_{BB} +c^4 f_{WW} + c^2 s^2 f_{BW}}{2 c^2} \label{eq:g} \;\; , \\
g^{(3)}_{H Z Z} &=& \left( \frac{ g M_W v^2}{\Lambda^2} \right) 
	              \frac{f_{\Phi,1}-f_{\Phi,2}}{4 c^2} \;\; , \nonumber \\
g^{(1)}_{H W W} &=& \left( \frac{g M_W}{\Lambda^2} \right) 
                      \frac{f_{W}}{2}  \;\; , \nonumber \\
g^{(2)}_{H W W} &=& - \left( \frac{g M_W }{\Lambda^2} \right) 
  f_{WW}  \;\; , \nonumber \\
g^{(3)}_{H W W} 
&=& -  \left( \frac{ g M_W v^2}{\Lambda^2} \right) 
	              \frac{f_{\Phi,1}+2 f_{\Phi,2}}{4}  \;\; ,
\nonumber
\end{eqnarray}
where $s$ and $c$ stand for the sine and cosine of the weak mixing
angle respectively.  We notice that we have rescaled the coefficient
$f_{GG}$ of the gluon-gluon operator in terms of a coupling $f_g$ also
including a loop suppression factor. In this way an anomalous
gluon-gluon coupling $f_g\sim {\mathcal O} (1-10)$ gives a
contribution comparable to the SM top loop. For the operators
involving electroweak gauge bosons we have kept the normalization
commonly used in the pre-LHC studies, for example, in
Refs.~\cite{Hagiwara:1993ck,Hagiwara:1993qt, Hagiwara:1995vp,
  Eboli:1999pt, GonzalezGarcia:1999fq}.  The couplings $g^{(3)}_{H Z
  Z}$ and $g^{(3)}_{H W W}$ include the effects arising from the
contribution of the operators ${\cal O}_{\Phi,1}$ and ${\cal
  O}_{\Phi,2}$ to the renormalization of the weak boson masses and the
Higgs field wave function.  \smallskip

For the sake of concreteness in this work we focus our attention on
modifications of the Higgs couplings to gauge bosons associated with
the five operators ${\cal O}_{GG}$, ${\cal O}_{BB}$, ${\cal O}_{WW}$,
${\cal O}_{B}$, and ${\cal O}_{W}$.  The operator ${\cal O}_{BW}$
contributes at tree level to the $W^3$--$B$ mixing and is therefore
very strongly constrained by the electroweak precision
data~\cite{DeRujula:1991se, Hagiwara:1993ck, Hagiwara:1993qt,
  Alam:1997nk}.  Similarly ${\cal O}_{\Phi, 1}$ contributes to the $Z$
mass but not to the $W$ mass and it is severely constrained by the
$\rho$ parameter.  Moreover the operators ${\cal O}_{\Phi, 1}$ and
${\cal O}_{\Phi, 2}$ lead to a multiplicative contribution to the SM
Higgs couplings to $ZZ$ and $WW$.  Thus in the present analysis we do
not consider effects associated with ${\cal O}_{BW}$, ${\cal O}_{\Phi,
  1}$ and ${\cal O}_{\Phi, 2}$ as their coefficients are already very
constrained or their possible effect on the measured Higgs observables
is degenerated with that of the five operators considered.  Their
impact on the Higgs phenomenology can be seen in
Refs.~\cite{cms8summer, Espinosa:2012ir, Ellis:2012rx, Carmi:2012yp,
  Azatov:2012bz, Klute:2012pu,Bonnet:2011yx}.  \smallskip

Notice also that one expects the contribution of new physics to the
five operators considered to take place at loop
level~\cite{Arzt:1994gp}.  Therefore, we expect that the largest
effect of these effective interactions should appear in the couplings
of the Higgs to photon--photon and gluon--gluon since these couplings
take place through loop effects in the SM. \smallskip

One important property of the operators ${\cal O}_B$ and ${\cal O}_W$
is that they also modify the triple gauge--boson couplings $\gamma W^+
W^-$ and $Z W^+ W^-$. Consequently they can be directly probed in
additional channels not directly involving the Higgs
boson~\cite{Eboli:1999pt,Aad:2012mr,Martelli:2012ea}.  The triple
gauge--boson effective interaction can be rewritten in the standard
parametrization of the $C$ and $P$ even interactions~\cite{Hagiwara:1986vm}:
\begin{equation}
{\cal L}_{WWV} = -i g_{WWV} \Bigg\{ 
g_1^V \Big( W^+_{\mu\nu} W^{- \, \mu} V^{\nu} 
  - W^+_{\mu} V_{\nu} W^{- \, \mu\nu} \Big) 
 +  \kappa_V W_\mu^+ W_\nu^- V^{\mu\nu}
+ \frac{\lambda_V}{m_W^2} W^+_{\mu\nu} W^{- \, \nu\rho} V_\rho^{\; \mu}
 \Bigg\}
\;\;,
\label{eq:classical}
\end{equation}
where $g_{WW\gamma} = e$ and $g_{WWZ} = e/(s\,c)$. In general these
vertices involve six dimensionless couplings $g_{1}^{V}$, $\kappa_V$,
and $\lambda_V$ $(V = \gamma$ or $Z)$.  Notwithstanding the
electromagnetic gauge invariance requires that $g_{1}^{\gamma} = 1$,
while the remaining five free couplings are related to the
dimension--six operators that we are considering:
\begin{eqnarray}
\Delta g_1^Z& = g_1^Z-1= &\frac{1}{2}\frac{m_Z^2}{\Lambda^2}f_W \;\;, 
\nonumber \\
\Delta \kappa_\gamma & = \kappa_\gamma -1 = 
&  \frac{1}{2}\frac{m_W^2}{\Lambda^2}
\Big(f_W + f_B\Big) \;\;, \label{eq:wwv}\\
\Delta \kappa_Z & = \kappa_Z -1 = &  \frac{1}{2}\frac{m_Z^2}{\Lambda^2}
  \Big(c^2 f_W - s^2 f_B\Big)\;\;, 
\nonumber \\
\lambda_\gamma &= \lambda_Z = & 
 0 \;\;. \nonumber
\end{eqnarray}

In summary, in the theoretical framework that we are using the
observables depend upon 5 parameters, $f_g$, $f_B$, $f_W$, $f_{BB}$ and
$f_{WW}$. In what follows for the sake of simplicity we focus on two
different scenarios:
\begin{itemize}

\item {\bf Scenario I:} we impose that $f_W=f_B$ and
  $f_{BB}=f_{WW}$. This scenario has three free parameters ($f_W$,
  $f_{WW}$ and $f_g$) and it exhibits a constraint between the three
  couplings of the Higgs to electroweak vector bosons. This scenario
  predicts the existence of anomalous triple electroweak gauge--boson
  interactions.

\item {\bf Scenario II:} we set $f_W = f_B = 0$ and
  $f_{WW}=f_{BB}$. This scenario has two free parameters ($f_g$ and
  $f_{BB}=f_{WW}$) and it can be considered the low--energy limit of
  an extension of the SM that contains an extra heavy scalar
  multiplet; for details see Ref.~\cite{Alam:1997nk}. Moreover, this
  scenario cannot be constrained by data on triple gauge--boson
  couplings.
\end{itemize}

The above relations (\ref{eq:wwv}) allow us to constrain the couplings
$f_B$ and $f_W$ using the available experimental bounds on the
effective couplings $\Delta g_1^Z$, $\Delta \kappa_\gamma$ and $\Delta
\kappa_Z$~\cite{Nakamura:2010zzi}.  Nevertheless these experimental
bounds are usually obtained assuming only one anomalous operator
different from the SM value at a time, an assumption which is not
consistent with our scenario I.  For this reason strictly speaking one
cannot apply the exclusion limits in Ref.~\cite{Nakamura:2010zzi} to
this scenario.  Nevertheless if we assume no strong cancellations
between the contributions of the different triple gauge--effective
operators we can estimate the size of the exclusion limits on $f_W$
and $f_B$. Using the 95\% CL regions from Ref.~\cite{Nakamura:2010zzi}
on $\Delta g_1^Z$, $\Delta \kappa_\gamma$ or $\Delta \kappa_Z$ we
obtain that in scenario I the 95\% CL regions on $f_W=f_B$ are
$[-13,7]$ TeV$^{-2}$, $[-18,9]$ TeV$^{-2}$, and $[-85,20]$ TeV$^{-2}$
respectively. Notice also that, LHC already with present runs has
potential to constraint the triple-gauge boson vertices
\cite{Eboli:2010qd} and the collaborations are starting to look for
deviations~\cite{Aad:2012mr, Martelli:2012ea}.  However, at present,
their individual limits have not reached the level of the LEP bounds
yet.

%%%%%%%%%%%%%%%%%%%%%%%%%%%%%%%%%%%%%%%%%%%%%%%%%%%%%%%%%%%%%%%%%%%%%%
\section{Analyses framework}

In order to obtain the present constraints on the Higgs anomalous
interactions we perform a chi--square test using the available data on
the signal strength ($\mu$) from Tevatron, LHC at 7 TeV and LHC at 8
TeV. We assume that the correlations between the different channels
are negligible except for the theoretical uncertainties which are
treated with the pull method~\cite{Fogli:2002pt,GonzalezGarcia:2007ib}
in order to account for their correlations.  \smallskip

Schematically we can write
\begin{equation}
\chi^2 =  \min_{\xi_{pull}}\sum_{j} 
\frac{(\mu_j - \mu_j^{\rm exp})^2}{\sigma_j^2}
+ \sum_{pull} \left ( \frac{\xi_{pull}}{\sigma_{pull}} \right )^2
\end{equation}
where $j$ stands for channels presented in Tables~\ref{tab:tev_lhc7}
and~\ref{tab:lhc8}.  We denote the theoretically expected signal as
$\mu_j$, the observed best fit values as $\mu_j^{\rm exp}$ and errors
as $\sigma^{+,-}_j$. As we can see from these tables the errors are
not symmetric, showing a deviation from a Gaussian behavior as
expected from the still low statistics. In our calculations we make
the errors in each channel symmetric by taking
\begin{equation}
     \sigma_j = \sqrt{ \frac{(\sigma_j^+)^2 + (\sigma_j^-)^2}{2}  } \;\; .
\end{equation}

Concerning the theoretical uncertainties, the largest are associated
with the gluon fusion subprocess and to account for these errors we
introduce two pull factors, one for the Tevatron ($\xi_{T}$) and one
for the LHC at 7 TeV and LHC at 8 TeV ($\xi_{L}$).  They modify the
corresponding predictions as shown in Eqs.~(\ref{eq:mu_vbf}) and
(\ref{eq:sigma_inclu}). We consider that the errors associated with
the pulls are $\sigma_T = 0.4$ and $\sigma_L = 0.15$.  As statistics
build up it will be necessary to introduce pulls associated with the
theoretical uncertainties for the other production mechanisms as well
as possible systematic correlated errors, however at this moment these
are sub--leading effects. \smallskip

%%%%%%%%%%%%%%%%%%%%%%%%%%%%%%%%%%%
\begin{table}
\begin{tabular}{|c|c|c|c|c|}
  \hline
  Channel & $\mu^{exp}$ & Comment
  \\
  \hline
  $p \bar{p} \rightarrow W^+W^-$ & $0.3^{+1.1}_{-0.3}$ & CDF \& D\O 
  ~\cite{tevanew} %1207.0449
  \\
  \hline
  $p \bar{p} \rightarrow b \bar{b}$  & $2.0^{+0.7}_{-0.7}$ & CDF \& D\O 
  ~\cite{tevanew} %1207.0449
  \\
  \hline
  $p \bar{p}\rightarrow \gamma \gamma$  & $3.6^{+3.0}_{-2.5}$ & CDF \& D\O
  ~\cite{tevanew} %1207.0449
  \\
  \hline
  $p p\rightarrow \tau \bar{\tau}$  & $0.2^{+1.7}_{-1.9}$ & ATLAS
  ~\cite{atlas7new}%1207.0319
  \\
  \hline
  $p p\rightarrow b \bar{b}$  & $0.5^{+2.1}_{-2.0}$ &  ATLAS
  ~\cite{atlas7new}%1207.0319
  \\
  \hline
  $p p\rightarrow Z Z^*\rightarrow \ell^+ \ell^- \ell^+ \ell^-$  & $1.4^{+1.3}_{-0.8}$ 
  &  ATLAS~\cite{atlas7new}%1207.0319
  \\
  \hline
  $p p\rightarrow W W^*\rightarrow \ell^+ \nu \ell^- \bar{\nu}$  & $0.5^{+0.6}_{-0.6}$ & ATLAS
  ~\cite{atlas7new}%1207.0319
  \\
  \hline
  $p p\rightarrow \gamma \gamma$  & $2.2^{+0.8}_{-0.8}$ &  ATLAS
~\cite{atlasconf12091}
  \\
  \hline
  $p p\rightarrow \tau \bar{\tau}$  & $0.6^{+1.1}_{-1.2}$ & CMS
  ~\cite{Chatrchyan:2012tx}  %1202.1488
  \\
  \hline
  $p p\rightarrow b \bar{b}$  & $0.5^{+1.1}_{-1.0}$ & CMS~\cite{cmspashig12020}
  \\
  \hline
  $p p\rightarrow Z Z^*\rightarrow \ell^+ \ell^- \ell^+ \ell^-$  & $0.6^{+0.9}_{-0.6}$ & CMS
  ~\cite{Chatrchyan:2012tx}  %1202.1488
  \\
  \hline
  $p p\rightarrow W W^*\rightarrow \ell^+ \nu \ell^- \bar{\nu}$  & $0.4^{+0.6}_{-0.6}$ & CMS
  ~\cite{Chatrchyan:2012tx}  %1202.1488
  \\
  \hline
 $p p\rightarrow \gamma \gamma$ Untagged 0 & $3.2^{+1.9}_{-1.8}$  & 
CMS~\cite{cmspashig12015}
  \\
  \hline
  $p p\rightarrow \gamma \gamma$ Untagged 1 & $0.7^{+0.9}_{-1.0}$  & 
CMS~\cite{cmspashig12015}
  \\
  \hline
  $p p\rightarrow \gamma \gamma$ Untagged 2 & $0.7^{+1.2}_{-1.1}$  & 
CMS~\cite{cmspashig12015}
  \\
  \hline
  $p p\rightarrow \gamma \gamma$ Untagged 3 & $1.5^{+1.6}_{-1.6}$  & 
CMS~\cite{cmspashig12015}
  \\
  \hline
 $p p\rightarrow \gamma \gamma j j$ & $4.2^{+2.0}_{-2.0}$  & 
CMS~\cite{cmspashig12015}
  \\
  \hline
\end{tabular}
\caption{Processes considered in our analyses for the LHC 7 TeV run and for 
  the Tevatron. We present  the  errors and best fit point for the signal 
  strength for each topology.}
\label{tab:tev_lhc7}
\end{table}
%%%%%%%%%%%%%%%%%%%%%%%%%%%%%%%%%%%

One important approximation in our analyses is that we neglect the
effects associated with the distortions of the kinematical
distributions of the final states due to the Higgs anomalous couplings
arising from their non SM-like Lorentz structure.  Thus we implicitly
assume that the anomalous contributions have the same detection
efficiencies as the SM Higgs. A full simulation of the Higgs anomalous
operators taking advantage of their special kinematical features would
increase the current sensitivity on the anomalous couplings.  It would
also allow for breaking degeneracies with those operators which only
lead to an overall modification of strength of the SM vertices. But at
present there is not enough public information to perform such
analysis outside of the experimental collaborations. \smallskip

%%%%%%%%%%%%%%%%%%%%%%%%%%%%%%%%%%%
\begin{table}
\begin{tabular}{|c|c|c|}
\hline
 Channel & $\mu^{exp}$ & Comment
\\
\hline
 $p p\rightarrow Z Z^*\rightarrow \ell^+ \ell^- \ell^+ \ell^-$ & 
  $1.3^{+0.6}_{-0.6}$ &  ATLAS @ 7 and 8 TeV ~\cite{atlas8summer}
\\
\hline
 $p p\rightarrow \gamma \gamma$ &  $1.8^{+0.6}_{-0.7}$ & ATLAS
~\cite{atlasconf12091}
\\
\hline
 $p p\rightarrow \tau \bar{\tau}$ &  $-0.2^{+0.8}_{-0.7}$   & CMS
@ 7 and 8 TeV~\cite{cms8summer} 
\\
\hline
 $p p\rightarrow b \bar{b}$ &  $0.4^{+0.9}_{-0.8}$  &   
CMS~\cite{cmspashig12020}
\\
\hline
 $p p\rightarrow Z Z^*\rightarrow \ell^+ \ell^- \ell^+ \ell^-$ & $0.7^{+0.6}_{-0.4}$ & 
CMS~\cite{cmspashig12020}
\\
\hline
 $p p\rightarrow W W^*\rightarrow \ell^+ \nu \ell^- \bar{\nu}$   &  $0.6^{+0.4}_{-0.4}$ & 
CMS @ 7 and 8 TeV~\cite{cms8summer} 
\\
\hline
 $p p\rightarrow \gamma \gamma$ Untagged 0 & $1.5^{+1.2}_{-1.2}$   & 
CMS~\cite{cmspashig12015}
\\
\hline
 $p p\rightarrow \gamma \gamma$ Untagged 1 & $1.5^{+1.0}_{-1.0}$  & 
CMS~\cite{cmspashig12015}
\\
\hline
 $p p\rightarrow \gamma \gamma$ Untagged 2 &  $1.0^{+1.1}_{-1.2}$  & 
CMS~\cite{cmspashig12015}
\\
\hline
 $p p\rightarrow \gamma \gamma$ Untagged 3 & $3.8^{+1.7}_{-1.8}$ & 
CMS~\cite{cmspashig12015}
\\
\hline
 $p p\rightarrow \gamma \gamma j j$  loose &  $-0.6^{+2.1}_{-2.0}$  & 
CMS~\cite{cmspashig12015}
\\
\hline
 $p p\rightarrow \gamma \gamma j j$  tight & $1.3^{+1.5}_{-1.6}$  & 
CMS~\cite{cmspashig12015}
\\
\hline
\end{tabular}
\caption{Available data including the 8 TeV run. We present
  the  errors and best fit point for the signal strength for each channel. 
  The data   that have been  combined  is indicated as  ``@ 7 and 8 TeV''.
}
\label{tab:lhc8}
\end{table}
%%%%%%%%%%%%%%%%%%%%%%%%%%%%%%%%%%%

In order to predict the modification of the observables we need to
include the effect of the anomalous operators in the production
channels as well as in the decay branching ratios.  As a first
approximation we can assume that the $K$ factor associated with higher
order corrections is the same for the SM and anomalous contributions,
so we write
\begin{equation}
\sigma^{ano}_Y = \left . \frac{\sigma^{ano}_{Y}}{\sigma^{SM}_{Y}}
             \right |_{tree} \; \left .\sigma^{SM}_{Y} \right |_{soa}
\label{eq:sigma_cor}
 \end{equation}
 where the ratio of the anomalous and SM cross sections of the
 subprocess $Y$ ($= gg$, VBF, $VH$ or $t\bar{t}H$) is evaluated at
 tree level and it is multiplied by the value for the
 state--of--the--art SM cross section calculations ($\sigma^{SM}_{Y}
 |_{soa}$) presented in Ref.~\cite{Dittmaier:2011ti}. Analogously we
 write the decay width into the final state $X$ as
\begin{equation}
\Gamma^{ano} (h \to X) = \left . \frac{\Gamma^{ano} (h\to X)}{\Gamma^{SM} (h \to X)}
             \right |_{tree} \; \left .\Gamma^{SM} (h \to X) \right |_{soa}
\label{eq:width_cor}
\end{equation}
where the SM result $\Gamma^{SM} (h \to X) |_{soa}$ is also obtained
from Ref.~\cite{Dittmaier:2011ti}.  The total width and branching
ratios are evaluated following this recipe.  We use the SM cross
sections and decay widths and compute our predictions for $m_H=125$
GeV.  The observed Higgs mass by ATLAS ($126.5$
GeV)~\cite{atlas8summer} and CMS ($125.3$ GeV)~\cite{cms8summer} are
compatible within the experimental errors. We verified that the impact
of changing the Higgs mass to $126.5$ GeV is a sub--leading effect and
does not alter our results.  We did not include in our analyses an
eventual invisible decay of the Higgs
~\cite{Espinosa:2012vu,Raidal:2011xk}, therefore the total width is
obtained by summing over the decays into the SM particles.  \smallskip

The search for Higgs decaying into $b \bar{b}$ pairs takes place
through Higgs production in association with a $W$ or a $Z$ so we can
write
\begin{equation}
\mu_{b \bar{b}} = \frac{\sigma^{ano}_{WH} + \sigma^{ano}_{ZH}}
                      {\sigma^{SM}_{WH} + \sigma^{SM}_{ZH}} ~\otimes~
                 \frac{ \hbox{Br}^{ano}[h \to b \bar{b}]}
                      { \hbox{Br}^{SM }[h \to b \bar{b}]}
\label{eq:mu_bbar}
\end{equation}
with the superscripts $ano$ ($SM$) standing for the value of the
observable considering the anomalous and SM interactions (pure SM
contributions).  \smallskip

The CMS analyses of the 7 (8) TeV data separate the $\gamma\gamma$
final into 5 (6) categories and the contribution of each production
mechanism to a given category is presented in Table 2 of
Ref.~\cite{cmspashig12015}. Therefore, we write the theoretical signal
strength in these cases as
\begin{equation}
\mu_{\gamma\gamma}^{CMS} = \frac {\epsilon_{gg}\sigma_{gg}^{ano} (1+\xi_g) + \epsilon_{VBF}\sigma^{ano}_{VBF} 
        + \epsilon_{VH} \left ( \sigma^{ano}_{WH} + \sigma^{ano}_{ZH} \right ) 
        + \epsilon_{t\bar{t}H}\sigma^{ano}_{t\bar{t}H}}
    {\epsilon_{gg}\sigma_{gg}^{SM} + \epsilon_{VBF}\sigma^{SM}_{VBF} + \epsilon_{VH}\left ( \sigma^{SM}_{WH}
      +\sigma^{SM}_{ZH} \right ) + \epsilon_{t\bar{t}H}\sigma^{SM}_{t\bar{t}H}} ~\otimes~
                 \frac{ \hbox{Br}^{ano}[h \to \gamma\gamma]}
                      { \hbox{Br}^{SM }[h \to \gamma\gamma]} \;\; ,
\label{eq:mu_vbf}
\end{equation}
where $\xi_g$ is the pull associated with the gluon fusion cross
section uncertainties, and the branching ratio and the anomalous cross
sections are evaluated using the prescriptions (\ref{eq:sigma_cor})
and (\ref{eq:width_cor}). The weight of the different channels to each
category is encoded in the parameters $\epsilon_{X}$ with $X = VBF$,
$gg$, $VH$, and $t\bar{t}H$ and they are presented in
Tables~\ref{tab:eps7} and~\ref{tab:eps8}. \smallskip

%%%%%%%%%%%%%%%%%%%%%%%%%%%%%%%%%%%

\begin{table}
\begin{tabular}{|c|c|c|c|c|}
\hline
 Channel & $\epsilon_{gg}$ & $\epsilon_{VBF}$ & $\epsilon_{VH}$ & $\epsilon_{t\bar{t}H}$
\\
\hline
 $p p\rightarrow \gamma \gamma$ Untagged 0 & 0.13 & 0.46 & 0.70 & 1
\\
\hline
 $p p\rightarrow \gamma \gamma$ Untagged 1 & 0.57 & 0.49 & 0.67 & 1
\\
\hline
 $p p\rightarrow \gamma \gamma$ Untagged 2 & 1 & 0.56 & 0.76 & 0
\\
\hline
 $p p\rightarrow \gamma \gamma$ Untagged 3 & 1 & 0.56 & 0.76 & 0
\\
\hline
 $p p\rightarrow \gamma \gamma j j$ & 0.029 & 1 & 0.019 & 0
\\
\hline
\end{tabular}
\caption{Weight of each production mechanism for the different $\gamma\gamma$ categories
in the CMS analyses of the 7 TeV data.}
\label{tab:eps7}
\end{table}
%%%%%%%%%%%%%%%%%%%%%%%%%%%%%%%%%%%

%%%%%%%%%%%%%%%%%%%%%%%%%%%%%%%%%%%
\begin{table}
\begin{tabular}{|c|c|c|c|c|}
\hline
 Channel & $\epsilon_{gg}$ & $\epsilon_{VBF}$ & $\epsilon_{VH}$ & $\epsilon_{t\bar{t}H}$
\\
\hline
 $p p\rightarrow \gamma \gamma$ Untagged 0 & 0.11 & 0.25 & 0.48 & 1
\\
\hline
 $p p\rightarrow \gamma \gamma$ Untagged 1 & 0.59 & 0.50 & 0.72 & 1
\\
\hline
 $p p\rightarrow \gamma \gamma$ Untagged 2 & 1 & 0.54 & 0.58 & 0
\\
\hline
 $p p\rightarrow \gamma \gamma$ Untagged 3 & 1 & 0.54 & 0.78 & 0
\\
\hline
 $p p\rightarrow \gamma \gamma j j$ loose & 0.094 & 1 & 0.064 & 0
\\
\hline
 $p p\rightarrow \gamma \gamma j j$ tight & 0.024 & 1 & 0 & 0
\\
\hline
\end{tabular}
\caption{Same as Table~\ref{tab:eps7} but for the 8 TeV CMS data.}
\label{tab:eps8}
\end{table}
%%%%%%%%%%%%%%%%%%%%%%%%%%%%%%%%%%%

With the exception of the above processes, all other channels are
treated as inclusive, so we write the expected signal strength of the
final state $F$ as
\begin{equation}
\mu_F = \frac {\sigma_{gg}^{ano} (1+\xi_g) + \sigma^{ano}_{VBF} 
                + \sigma^{ano}_{WH} + \sigma^{ano}_{ZH}
		+ \sigma_{t\bar{t}H}^{ano}}
              {\sigma_{gg}^{SM} + \sigma^{SM}_{VBF} 
              + \sigma^{SM}_{WH} + \sigma^{SM}_{ZH}
              + \sigma_{t\bar{t}H}^{SM}
              } ~\otimes~
                 \frac{ \hbox{Br}^{ano}[h \to F]}
                      { \hbox{Br}^{SM }[h \to F]} \;\; .
\label{eq:sigma_inclu}
\end{equation}
Here we also use Eqs.~(\ref{eq:sigma_cor}) and (\ref{eq:width_cor}) to
obtain the anomalous cross sections and branching ratios.  \smallskip

For some final states the available LHC 8 TeV data has been presented
combined with the 7 TeV results. Given the limited available
information on errors and correlations, we construct the expected
theoretical signal strength as an average of the expected signal
strengths for the center--of--mass energies of 7 and 8 TeV. We weight
the contributions by the total number of events expected at each
energy in the framework of the SM, \ie\ given a final state $X$ we
evaluate
\begin{equation}
\mu_X^{comb}  =\frac{    \mu_X^{7 {\rm TeV}} ~ \sigma_{X}^{SM,7 {\rm TeV} } 
{\cal L}_{7 {\rm TeV}}  
                    +  \mu_X^{8 {\rm TeV}} ~ 
\sigma_{X}^{SM,8 {\rm TeV}} {\cal L}_{8 {\rm TeV}}  }
  { \sigma_X^{SM, 7 {\rm TeV}} {\cal L}_{7 {\rm TeV}}  + 
\sigma_X^{SM,8 {\rm TeV}} 
               {\cal L}_{8 {\rm TeV}}  }  \;\; ,
\end{equation}
where $ {\cal L}_{7 (8) {\rm TeV}}$ stands for the integrated
luminosity at 7 (8) TeV accumulated in the channel being analyzed.
When considering the full available data set we consider all the
processes in Table~\ref{tab:tev_lhc7} and \ref{tab:lhc8}, neglecting
the LHC 7 TeV processes whose data has been combined with the 8 TeV
run; we indicate in Table~\ref{tab:lhc8} that the data has been
combined by ``@ 7 and 8 TeV''.\smallskip

The evaluation of the relevant tree level cross sections was done
using the package MadGraph5~\cite{Alwall:2011uj} with the anomalous
Higgs interactions introduced using
FeynRules~\cite{Christensen:2008py}.  We also cross checked our
results using COMPHEP~\cite{Pukhov:1999gg, Boos:2004kh} and
VBFNLO~\cite{Arnold:2011wj}. The evaluation of the partial width was
done using the expressions presented in
Ref.~\cite{GonzalezGarcia:1999fq}. \smallskip

%%%%%%%%%%%%%%%%%%%%%%%%%%%%%%%%%%%%%%%%%%%%%%%%%%%%%%%%%%%%%%%%%%%%%%
\section{Constraints on the Higgs anomalous interactions }

\begin{figure}[h!]
  \centering
 \includegraphics[width=0.85\textwidth]{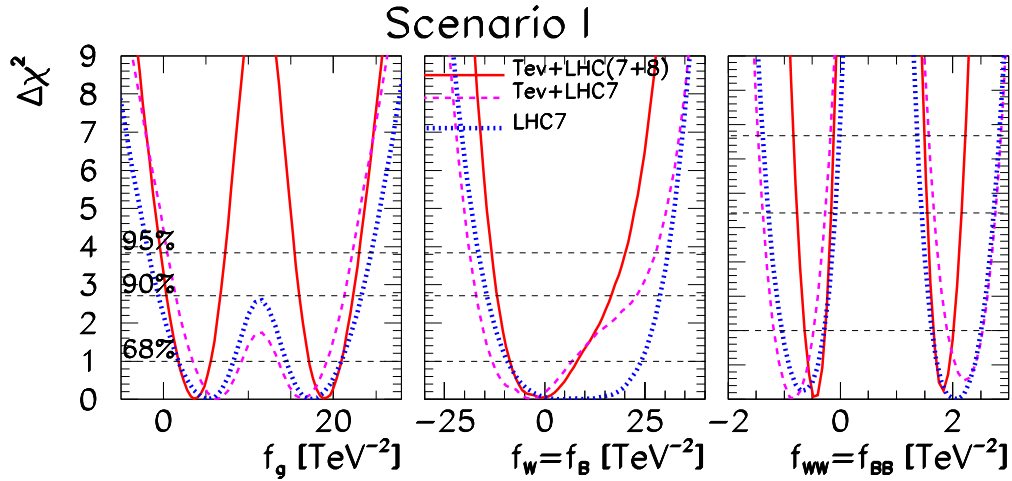}
 \caption{ The left (central, right) panel exhibits $\Delta \chi^2 $
   as a function of $f_{g}$ ($f_{W}$, $f_{WW}$) in the framework of
   scenario I. Each panel contains three lines: the dotted (dashed)
   line was obtained using only the LHC 7 TeV (LHC 7 TeV and Tevatron)
   data while the solid line stands for the result using all the
   available data. In each panel $\Delta \chi^2$ is marginalized over
   the two undisplayed parameters.}
\label{fig:chi2_I}
\end{figure}

\begin{figure}[h!]
  \centering
 \includegraphics[width=0.6\textwidth]{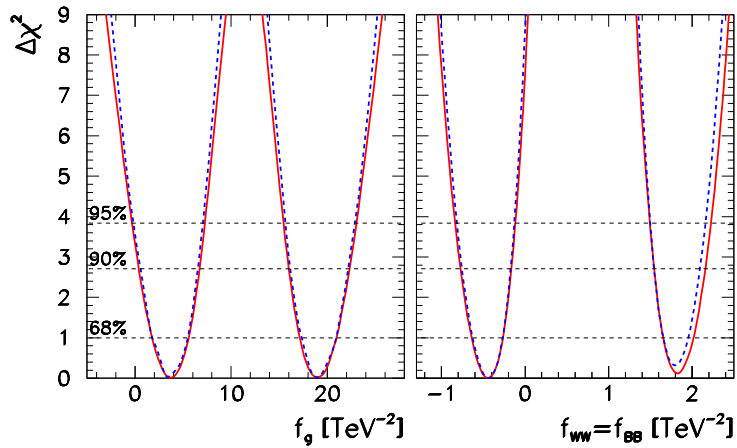}
 \caption{$\Delta \chi^2$ as a function of $f_{g}$ (left panel) and
   $f_{WW}=f_{BB}$ (right panel) for the full combined analysis. The
   solid lines correspond to scenario I in which $\Delta\chi^2$ is
   marginalized over the two undisplayed parameters in each panel:
   $f_{WW}=f_{BB}$ and $f_{W}=f_{B}$ in the left panel, and $f_{g}$
   and $f_{W}=f_{B}$ in the right panel.  The dashed lines correspond
   to scenario II, \ie\ imposing first the prior $f_{W}=f_{B}=0$ and
   then marginalizing over $f_{WW}=f_{BB}$ (left) and $f_g$ (right).}
\label{fig:chi2_II}
\end{figure}

\begin{figure}[h!]
  \centering
 \includegraphics[width=0.75\textwidth]{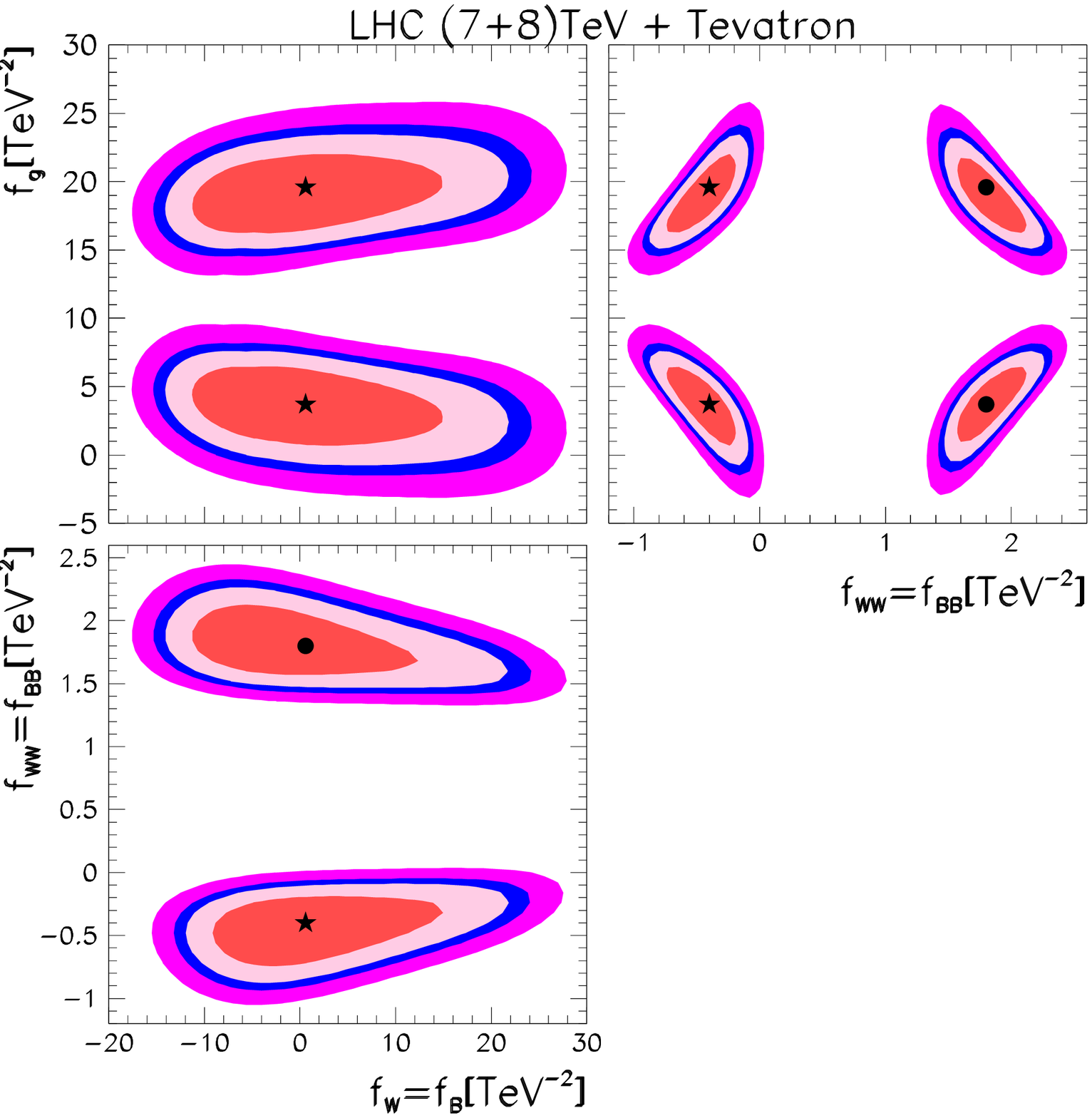}
 \caption{68\%, 90\%, 95\%, and 99\% CL (2dof) allowed regions of the
   plane $f_{WW} \otimes f_g$ (upper right panel), $f_{W} \otimes f_g$
   (upper left panel) and $f_{W} \otimes f_{WW}$ (lower panel) using
   all available data.  These results are obtained for scenario
   I and after marginalization over the undisplayed parameter in each
   panel.  The best fit points are indicated by a star while the second
local minima are indicated with a dot.}
\label{fig:correl_I}
\end{figure}

We next derive the allowed values of the Higgs interactions to vector
bosons using the available Tevatron data~\cite{tevanew}, ATLAS 7
TeV~\cite{atlas7new,atlasconf12091} and 8
TeV~\cite{atlas8summer,atlasconf12091} results, and CMS 7
TeV~\cite{Chatrchyan:2012tx, cmspashig12015, cmspashig12020} and 8 TeV
~\cite{cms8summer,cmspashig12015,cmspashig12020} data.  The results
are presented in Fig.~\ref{fig:chi2_I}--Fig.~\ref{fig:correl_II} where
several one-dimensional and two-dimensional projections of the
$\Delta\chi^2=\chi^2-\chi^2_{min}$ function(s) are shown.  We find
$\chi^2_{min}= 12.12$ $(12.13)$ in scenario I (II) for the global
analysis (\ie\ for a total number of 26 data points). The SM lays at
$\chi^2_{SM}=20.87$ \ie\ within the 96.7\% (98.7\%) CL region in the
three-- (two--)dimensional parameter space.  The corresponding best
fit values and 95\% allowed ranges are summarized in
Table~\ref{tab:bf95}.  \smallskip

Figure~\ref{fig:chi2_I} shows $\Delta \chi^2$ as a function of each of
the operator coefficients in scenario I after marginalizing with
respect to the two undisplayed ones. To illustrate the effect of the
different data sets, the results are shown for three combinations of
the available data: the dotted (dashed) line stands for the results
obtained using only the LHC 7 TeV (LHC 7 TeV and Tevatron) data while
the solid line is derived using the full available data set.  The
central panel of this figure displays the $\Delta \chi^2$ dependence
on $f_W$. As we can see the analysis of the LHC 7 TeV data only leads
to a large flat region around the minimum indicating that this data
set has a small sensitivity to $f_W$, \ie\ the Higgs couplings to $W$
and $Z$ pairs. This is expected since the $\gamma \gamma$ channel is
the dominant observable in this sample. The addition of the Tevatron
data, dominated by the Higgs associated production, enhances the
sensitivity to deviations in $HZZ$ and $HW^+W^-$, that is, to smaller
values of $f_W$.  The addition of the LHC 8 TeV results further
tightens the allowed values giving, for the global analysis, the
constraint $-13 \leq f_W\leq 20$ at 95\% CL. \smallskip

$\Delta \chi^2$ as a function of $f_g$ is shown in the left panel of
Figure~\ref{fig:chi2_I} where we see that the analysis present two
totally degenerate minima leading to two distinct allowed ranges.
This degeneracy -- as others that we encounter in this work -- is due
to the interference between the SM and anomalous contributions.  We
see that before the inclusion of LHC 8 TeV data, the two allowed
ranges overlapped at CL higher than 90 \%, while in the global
analysis they are separated at more than 3$\sigma$.  The value of the
gluon fusion cross section at the minima is around 43\% of its SM
value (see left panel of Fig.~\ref{fig:chi2_III}) and this cross
section is { highly} suppressed in the region between the minima. This
{highly} suppressed gluon fusion cross section was not completely
disfavored before the 8 TeV data because the CDF \& D\O\ and CMS @ 7
TeV $b\bar{b}$ channels still allowed for a larger $f_W$ coupling to
enhance associated production, compensating the large reduction of the
gluon fusion cross section (see also discussion of
Fig.~\ref{fig:correl_I}).  After the inclusion of the LHC 8 TeV data this
is no longer possible.  So altogether the global analysis constrains
$f_g$ to lie in one of the two intervals $[ -0.3 \;,\; 7.3]$ or $[15
\;,\; 23]$ at 95\% CL. \smallskip

The $\Delta \chi^2 $ dependence on $f_{WW}$ in the scenario I is
presented in the right panel of Fig.~\ref{fig:chi2_I}.  A salient
feature of this plot is that $\Delta \chi^2$ is concentrated around
two narrow non--overlapping regions centered around almost (but not
totally) degenerate minima. Unlike for $f_g$, the two minima in
$f_{WW}=f_{BB}$ are not fully degenerated because these operators
modify not only the Higgs coupling to photons but also to $WW$ and
$ZZ$ and the contributions to these last two vertices are slightly
different at the two minima.  Moreover, we also see that Tevatron data
has a limited impact on this parameter while the inclusion of the LHC
8 TeV results tighten the bounds on $f_{WW}$ which at 95\% CL is
bounded to lie in one of the two intervals $[ -0.8 \;,\; -0.1]$ or $[
1.5 \;,\; 2.2 ]$.  \smallskip

%%%%%%%%%%%%%%%%%%%%%%%%%%%%%%%%%%%%%%%%%%%%%%%%%%%%%
%%%%%%%%%%%%%%%%%%%%%%%%%%%%%%%%%%%%%%%%%%%%%%%%%%%%%

The dependence on the scenario considered is illustrated in 
Figure~\ref{fig:chi2_II} where we plot the $\Delta \chi^2$ dependence
on $f_g$ and $f_{WW}$ of the global analysis in scenarios I and II.
As we can see the results for both scenarios are almost coincident in both
panels. This is due to the fact that in scenario I the full available
data set is well described by $f_W = f_B \simeq 0$ for all allowed
values of $f_{WW}$ and $f_g$, consequently the two scenarios give very
similar results.  
\smallskip

%%%%%%%%%%%%%%%%%%%%%%%%%%%%%%%%%%%%%%%%%%%%%%%%%%%%%
%%%%%%%%%%%%%%%%%%%%%%%%%%%%%%%%%%%%%%%%%%%%%%%%%%%%%

Let us turn our attention towards the correlations between the three
free parameters of scenario I.  Figure~\ref{fig:correl_I} depicts
68\%, 90\%, 95\%, and 99\% CL (2dof) allowed regions of the $f_{WW} \otimes
f_g$ (upper right panel), $f_{W} \otimes f_g$ (upper left panel) and
$f_{W} \otimes f_{WW}$ (lower panel) planes using all attainable
data. We obtained these plots marginalizing over the free parameter
not appearing in each of the panels.

%%%%%%%%%%%%%%%%%%%%%%%%%%%%%%%%%%%%%%%%%%%%%%%%%%%%%
%%%%%%%%%%%%%%%%%%%%%%%%%%%%%%%%%%%%%%%%%%%%%%%%%%%%%

We can see from the upper right panel of Fig.~\ref{fig:correl_I} that
there are four well isolated allowed ``islands'' in the $f_{WW}
\otimes f_g$ plane. Moreover, within each of these islands $f_{WW}$
and $f_g$ are strongly correlated or anti--correlated. As mentioned
before, the existence of degenerate islands is due to the interference
between the SM and anomalous contributions which allow two different
values of the anomalous couplings to lead to the same cross section or
branching ratio. In the case at hand, the gluon fusion cross section
preferred by the fit is around 43\% of its SM value.  It is interesting 
to notice that  if the results from  the $b\bar{b}$ channel are removed from
the fit the vertical gap between the two islands on the left (or on the
right) disappears -- that is, intermediate values of $f_g$ 
(which correspond to further suppressed gluon fusion production) 
become allowed. This happens because the $b \bar{b}$ data, which is 
dominated by
associated production, constrains the coupling of the Higgs to $W$ and
$Z$ pairs.  In our framework, this leads to (a) an associated upper bound 
on the $H\gamma\gamma$ branching ratio, and (b) an upper bound on VBF
and associated production. 
$\gamma\gamma$ data mainly restricts the product of the gluon fusion cross
section and the Higgs branching ratio into photons, thus weakening the 
upper bound on the latter allows the former to have smaller values. 
Furthermore even smaller gluon fusion cross sections are permited
because of the possible increase in the VBF and associated production 
processes.

The upper left panel of Figure~\ref{fig:correl_I} shows the presence
of two isolated regions in the $f_W \otimes f_G$ plane and that there
is a very weak correlation between the parameters within each region. Here
again, the removal of the $b \bar{b}$ data leads to the disappearance
of the gap between the allowed regions.
The lower panel displays a behavior similar to the one observed in the
upper left, but in the $f_W \otimes f_{WW}$ plane.

For the sake of completeness we also show the results of the global analysis 
in scenario I in terms of the allowed ranges of Higgs production cross 
sections and  decay branching ratios in Fig.~\ref{fig:chi2_III} and 
Fig.~\ref{fig:correl_II}. The results shown in these figures 
are obtained by projecting the three-dimensional $\Delta\chi^2$ function 
on the displayed  observables and marginalizing on the independent undisplayed 
combination(s). 

Finally  we also verified that the results do
not change significantly when we do not employ the pulls to perform
the fit. This behavior could be anticipated since the experimental
errors are still much larger than the errors described by the pulls; a
situation that will change as more statistics accumulate. \smallskip

%%%%%%%%%%%%%%%%%%%%%%%%%%%%%%%%%%%%%%%%%%%%%%%%%%%%%%%%%%%%%%%%%%%%%%
\section{Discussion}

Once a Higgs boson like state has been discovered we must study its
properties to establish if it is indeed the state predicted by the
SM. In addition to that, it is also important to look for additional
states that might play a role in the electroweak symmetry breaking.
In this article we have studied the Higgs couplings to gauge bosons
using a model--independent characterization of the deviations with
respect to the SM values in terms of dimension--six operators and the
available data from Tevatron and LHC at 7 TeV and 8 TeV. This approach
still assumes that the Higgs field is a doublet of the $SU(2)_L$
symmetry and that the deviations of its couplings from the SM values
are due to additional heavy states. Notwithstanding, our framework
allows for independent modifications of the couplings to gluons,
photons and weak gauge bosons. \smallskip

%%%%%%%%%%%%%%%%%%%%%%%%%%%%%%%%%%%%%%%%%%%%%%%%%%%%%
\begin{figure}[h!]
  \centering
 \includegraphics[width=0.6\textwidth]{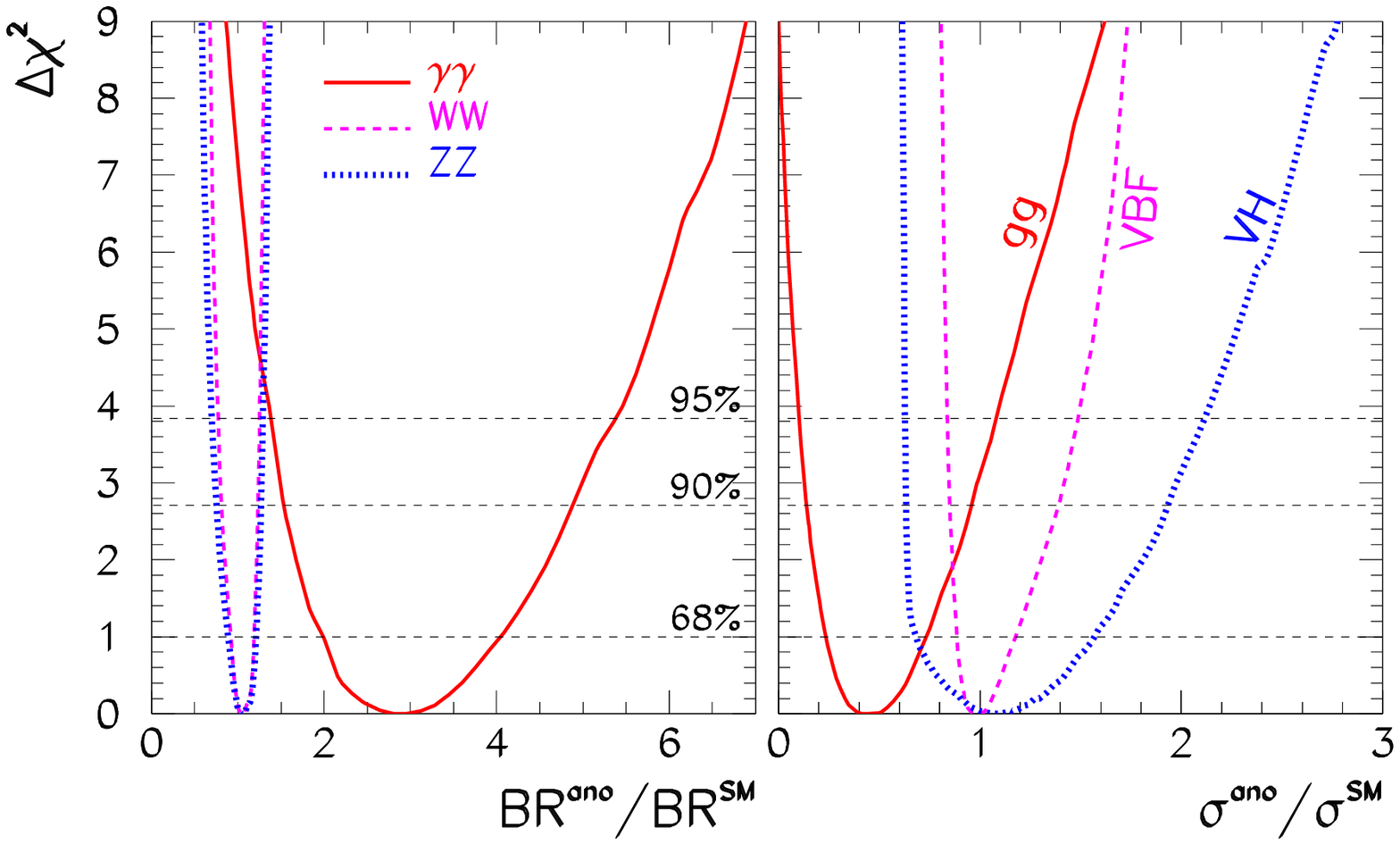}
 \caption{$\Delta \chi^2$ as a function of Higgs branching ratios into
   electroweak gauge bosons (left panel) and the 
   cross section for different production processes (right panel) 
   normalized to the
   SM values. In the left panel the solid (dashed, dotted) line stands
   for the branching ratio into $\gamma\gamma$ ($W^+ W^-$, $ZZ$),
   while, in the right panel, the solid (dashed, dotted) line
   represents the gluon fusion (VBF, VH) production cross section.}
\label{fig:chi2_III}
\end{figure}
%%%%%%%%%%%%%%%%%%%%%%%%%%%%%%%%%%%%%%%%%%%%%%%%%%%%%

In this study we have demonstrated that the present available data is
enough to start gaining some information on the different Higgs
couplings to gauge bosons.  For instance, our analyses indicate that a
reduced gluon fusion cross section is preferred when we use the full
available data set, with the most favored value being 43\% of the SM
value.  We can see this preference for a reduced gluon fusion cross
section in the right panel of Figure~\ref{fig:chi2_III} while the VBF
and associated production cross sections are in agreement with the SM
prediction. From this panel we can extract that the 95\% CL allowed
region of the gluon fusion cross section is $[0.1, 1.1]$ times the
corresponding SM value. This is consistent with the CMS
analyses~\cite{cms8summer} which, using a different framework, also
points in this direction as a reduced coupling of the Higgs to top
quarks is preferred by their results. \smallskip

Taking into account that the presently measured $\gamma\gamma$ yield
is above the SM prediction, the diminished gluon fusion cross section
points to an enhanced Higgs branching ratio in $\gamma\gamma$; a fact
that can be observed in our analyses. The left panel of
Figure~\ref{fig:chi2_III} shows that the $\gamma\gamma$ branching
ratio is indeed augmented, with a best fit value of 2.9 times the SM
value and the 95\% CL allowed region being $[1.4, 5.4]$ times the SM
branching ratio.  Furthermore, we can see from this panel that the Higgs
branching ratio into $W^+W^-$ and $ZZ$ is in agreement with the SM
expectations. \smallskip

\begin{figure}[h!]
  \centering
 \includegraphics[width=0.6\textwidth]{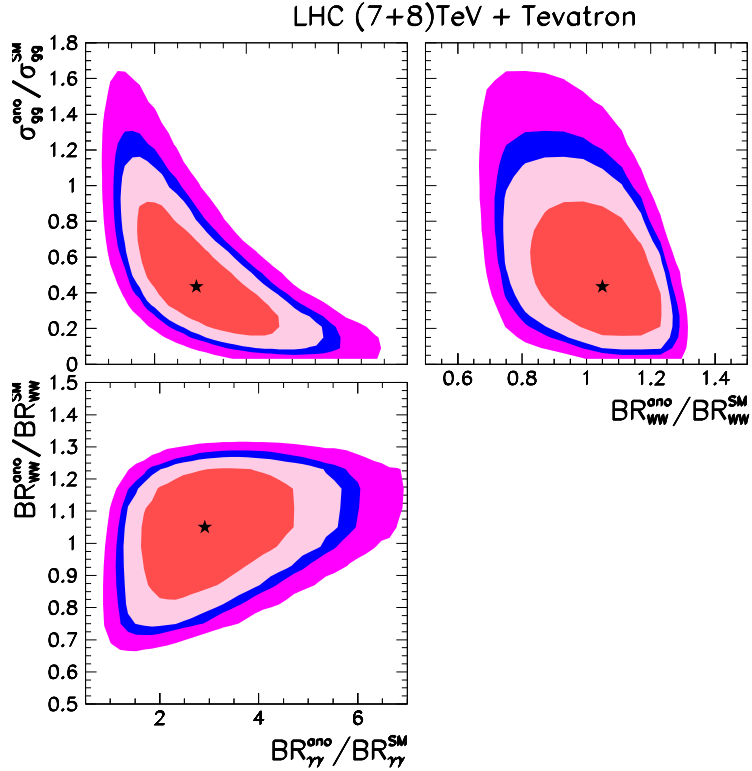}
 \caption{Allowed regions for several combinations of 
Higgs branching ratios and production cross section.
In each panel $\Delta\chi^2$ is marginalized with respect
to the combination of couplings independent of the two displayed
observables. 
As in Figure~\ref{fig:correl_I} the regions are shown at 
68\%, 90\%, 95\%, and 99\% CL (2dof).}  
\label{fig:correl_II}
\end{figure}

Presently the $\gamma \gamma$ channel is the best measured channel and
its rate is above the SM prediction. The operators ${\cal O}_{WW}$,
${\cal O}_{BB}$ and ${\cal O}_{GG}$ are the ones affecting this
channel, therefore they are the ones showing the largest impact of
the full data set.  This can be seen from the strong correlations and
the well isolated islands present in the upper right panel of
Figure~\ref{fig:correl_I} as well as by the correlations between the gluon 
fusion cross section and the Higgs branching ratios into electroweak gauge
bosons in Figure~\ref{fig:correl_II}.  
From the upper left panel of
this figure we can see an anti--correlation between the gluon fusion
cross section and the Higgs branching ratio into two photons; once
again it is clear that there is a preference for reduced gluon fusion cross
sections and enhanced decay into photon pairs. The other two panels 
of Figure~\ref{fig:correl_II} show the mild dependence 
of the Higgs branching ratio into $W^+W^-$ with the gluon fusion
cross section or the two photon branching ratio. \smallskip

%%%%%%%%%%%%%%%%%%%%%%%%%%%%%%%%%%%%%%%%%%%%%%%%%%%%%
%%%%%%%%%%%%%%%%%%%%%%%%%%%%%%%%%%%%%%%%%%%%%%%%%%%%%

Our analyses of scenario I also shows that the presently available data
prefers small values of $f_W= f_B$; see the central panel of Figure~\ref{fig:chi2_I}.
This indicates that large deviations in $HZZ$ and
$HW^+W^-$ interactions, as well as to triple gauge--boson couplings,
are not favoured.  This behavior was expected because the data points
for Higgs couplings to $W$'s and $Z$'s are in agreement with the SM within
$1\sigma$; see Figure~\ref{fig:chi2_III} left panel.  Furthermore, the
present direct constraints on triple gauge--boson vertices lead to
bounds on $f_W$ that are of the same order as the ones derived here
from Higgs phenomenology. So in the future the combined analysis of
Higgs data and measurements of the anomalous triple gauge--boson
couplings can be used to reduce the degeneracies observed in our
results since they present a different dependence on the anomalous
couplings $f_W$ and $f_B$; see Eqs.~(\ref{eq:g}) and~(\ref{eq:wwv}).
In this respect, it is interesting to notice that electroweak
precision measurements still give rise to the tightest limits on the
Higgs anomalous interactions~\cite{ Alam:1997nk,
  Hagiwara:1995vp}.\smallskip

%%%%%%%%%%%%%%%%%%%%%%%%%%%%%%%%%%%
\begin{table}
\begin{tabular}{|c|c|c|}
\cline{2-3}
 \multicolumn{1}{c|}{} & Best fit & 95\% CL allowed range
\\
\hline
 $f_W=f_B$ \footnotesize{(TeV$^{-2}$)} & -0.8 & $[-13, 20]$
\\
\hline
$f_{WW}=f_{BB}$ \footnotesize{(TeV$^{-2}$)} & -0.4, (1.8)

 & $[-0.8, -0.1]$ and $[1.5, 2.2]$
\\
\hline
$f_g$ \footnotesize{(TeV$^{-2}$)} & 3.7, 19  & $[-0.3, 7.3]$ and $[15, 23]$
\\
\hline
$BR^{ano}_{\gamma\gamma}/BR^{SM}_{\gamma\gamma}$ & 2.9 & $[1.4, 5.4]$
\\
\hline
$BR^{ano}_{WW}/BR^{SM}_{WW}$ & 1.1 & $[0.8, 1.3]$
\\
\hline
$BR^{ano}_{ZZ}/BR^{SM}_{ZZ}$ & 1.1 & $[0.7, 1.3]$
\\
\hline
$\sigma^{ano}_{gg}/\sigma^{SM}_{gg}$ & 0.4 & $[0.1, 1.1]$
\\
\hline
$\sigma^{ano}_{VBF}/\sigma^{SM}_{VBF}$ & 1.0 & $[0.8, 1.5]$
\\
\hline
$\sigma^{ano}_{VH}/\sigma^{SM}_{VH}$ & 1.1 & $[0.6, 2.1]$
\\
\hline
\end{tabular}
\caption{Best fit values and 95\% CL allowed ranges for the combination 
of all available data. For $f_g$ we show the two degenerate best fit values.
For $f_{WW}=f_{BB}$ together with the best fit we show in parenthesis the 
value at the second  minimum.} 
\label{tab:bf95}
\end{table}
%%%%%%%%%%%%%%%%%%%%%%%%%%%%%%%%%%%

We finish with a word of warning.  The precise numerical results presented 
here, that are summarized in Table~\ref{tab:bf95}, should be taken with a grain
of salt; due to the simplifying hypothesis used in our analyses we
should be aware that details can change if a more complete approach is
used. Nevertheless we verified that our results are rather robust when
we use only parts of the available data.  \smallskip

%%%%%%%%%%%%%%%%%%%%%%%%%%%%%%%%%%%%%%%%%%%%%%%%%%%%%%%%%%%%%%%%%%%%%%
\section*{Acknowledgments}

O.J.P.E is grateful to the Institute de Physique Th\'eorique de Saclay
for its hospitality.
O.J.P.E. is supported in part by Conselho
Nacional de Desenvolvimento Cient\'{\i}fico e Tecnol\'ogico (CNPq) and
by Funda\c{c}\~ao de Amparo \`a Pesquisa do Estado de S\~ao Paulo
(FAPESP); M.C.G-G is also supported by USA-NSF grant PHY-0653342, by
CUR Generalitat de Catalunya grant 2009SGR502 and together with J.G-F
by MICINN FPA2010-20807 and by consolider-ingenio 2010 program
CSD-2008-0037. J.G-F is further supported by Spanish ME FPU grant
AP2009-2546. T.C is supported by USA-NSF grant PHY-0653342.
We also acknowledge support of
EU grant FP7 ITN INVISIBLES (Marie Curie Actions PITN-GA-2011-289442).
O.J.P.E is grateful to R. Zukanovich Funchal for enlightening discussions.

%%%%%%%%%%%%%%%%%%%%%%%%%%%%%%%%%%%%%%%%%%%%%%%%%%%%%%%%%%%%%%%%%%%%%%

\bibliographystyle{h-physrev5}
\bibliography{referencias}

\begin{thebibliography}{10}

\bibitem{Englert:1964et}
F.~Englert and R.~Brout,
\newblock Phys.Rev.Lett. {\bf 13}, 321 (1964).
%%CITATION = PRLTA,13,321;%%

\bibitem{Higgs:1964pj}
P.~W. Higgs,
\newblock Phys.Rev.Lett. {\bf 13}, 508 (1964).
%%CITATION = PRLTA,13,508;%%

\bibitem{Higgs:1964ia}
P.~W. Higgs,
\newblock Phys.Lett. {\bf 12}, 132 (1964).
%%CITATION = PHLTA,12,132;%%

\bibitem{Guralnik:1964eu}
G.~Guralnik, C.~Hagen, and T.~Kibble,
\newblock Phys.Rev.Lett. {\bf 13}, 585 (1964).
%%CITATION = PRLTA,13,585;%%

\bibitem{Higgs:1966ev}
P.~W. Higgs,
\newblock Phys.Rev. {\bf 145}, 1156 (1966).
%%CITATION = PHRVA,145,1156;%%

\bibitem{Kibble:1967sv}
T.~Kibble,
\newblock Phys.Rev. {\bf 155}, 1554 (1967).
%%CITATION = PHRVA,155,1554;%%

\bibitem{atlas8summer}
ATLAS Collaboration, F.~Gianotti,
\newblock http://indico.cern.ch/conferenceDisplay.py?confId=197461.

\bibitem{cms8summer}
CMS Collaboration, J.~Incandela,
\newblock http://indico.cern.ch/conferenceDisplay.py?confId=197461.

\bibitem{Altarelli:2012dq}
For a recent review see, G.~Altarelli,
\newblock (2012), arXiv:1206.1476.
%%CITATION = ARXIV:1206.1476;%%

\bibitem{Dimopoulos:1979es}
S.~Dimopoulos and L.~Susskind,
\newblock Nucl.Phys. {\bf B155}, 237 (1979).
%%CITATION = NUPHA,B155,237;%%

\bibitem{Weinberg:1979bn}
S.~Weinberg,
\newblock Phys.Rev. {\bf D19}, 1277 (1979).
%%CITATION = PHRVA,D19,1277;%%

\bibitem{Hill:2002ap}
For a review see, C.~T. Hill and E.~H. Simmons,
\newblock Phys.Rept. {\bf 381}, 235 (2003), arXiv:hep-ph/0203079.
%%CITATION = HEP-PH/0203079;%%

\bibitem{Kaplan:1983fs}
D.~B. Kaplan and H.~Georgi,
\newblock Phys.Lett. {\bf B136}, 183 (1984).
%%CITATION = PHLTA,B136,183;%%

\bibitem{Kaplan:1983sm}
D.~B. Kaplan, H.~Georgi, and S.~Dimopoulos,
\newblock Phys.Lett. {\bf B136}, 187 (1984).
%%CITATION = PHLTA,B136,187;%%

\bibitem{Banks:1984gj}
T.~Banks,
\newblock Nucl.Phys. {\bf B243}, 125 (1984).
%%CITATION = NUPHA,B243,125;%%

\bibitem{Dugan:1984hq}
M.~J. Dugan, H.~Georgi, and D.~B. Kaplan,
\newblock Nucl.Phys. {\bf B254}, 299 (1985).
%%CITATION = NUPHA,B254,299;%%

\bibitem{Georgi:1984ef}
H.~Georgi, D.~B. Kaplan, and P.~Galison,
\newblock Phys.Lett. {\bf B143}, 152 (1984).
%%CITATION = PHLTA,B143,152;%%

\bibitem{Agashe:2004rs}
K.~Agashe, R.~Contino, and A.~Pomarol,
\newblock Nucl.Phys. {\bf B719}, 165 (2005), arXiv:hep-ph/0412089.
%%CITATION = HEP-PH/0412089;%%

\bibitem{Giudice:2007fh}
G.~Giudice, C.~Grojean, A.~Pomarol, and R.~Rattazzi,
\newblock JHEP {\bf 0706}, 045 (2007), arXiv:hep-ph/0703164.
%%CITATION = HEP-PH/0703164;%%

\bibitem{Buchmuller:1985jz}
W.~Buchmuller and D.~Wyler,
\newblock Nucl.Phys. {\bf B268}, 621 (1986).
%%CITATION = NUPHA,B268,621;%%

\bibitem{Leung:1984ni}
C.~N. Leung, S.~Love, and S.~Rao,
\newblock Z.Phys. {\bf C31}, 433 (1986).
%%CITATION = ZEPYA,C31,433;%%

\bibitem{tevanew}
The CDF Collaboration, the D0 Collaboration, Higgs Working Group, t.~T.~N.
  Physics,
\newblock (2012), arXiv:1207.0449.
%%CITATION = ARXIV:1207.0449;%%

\bibitem{atlas7new}
ATLAS Collaboration, G.~Aad {\em et~al.},
\newblock (2012), arXiv:1207.0319.
%%CITATION = ARXIV:1207.0319;%%

\bibitem{Chatrchyan:2012tx}
CMS Collaboration, S.~Chatrchyan {\em et~al.},
\newblock Phys.Lett. {\bf B710}, 26 (2012), arXiv:1202.1488.
%%CITATION = ARXIV:1202.1488;%%

\bibitem{cmspashig12015}
CMS Collaboration,
\newblock CMS PAS HIG-12-015.

\bibitem{cmspashig12020}
CMS Collaboration,
\newblock CMS PAS HIG-12-020.

\bibitem{atlasconf12091}
ATLAS Collaboration,
\newblock ATLAS-CONF-2012-091.

\bibitem{Hagiwara:1993ck}
K.~Hagiwara, S.~Ishihara, R.~Szalapski, and D.~Zeppenfeld,
\newblock Phys.Rev. {\bf D48}, 2182 (1993).
%%CITATION = PHRVA,D48,2182;%%

\bibitem{Hagiwara:1993qt}
K.~Hagiwara, R.~Szalapski, and D.~Zeppenfeld,
\newblock Phys.Lett. {\bf B318}, 155 (1993), arXiv:hep-ph/9308347.
%%CITATION = HEP-PH/9308347;%%

\bibitem{Hagiwara:1995vp}
K.~Hagiwara, S.~Matsumoto, and R.~Szalapski,
\newblock Phys.Lett. {\bf B357}, 411 (1995), arXiv:hep-ph/9505322.
%%CITATION = HEP-PH/9505322;%%

\bibitem{Eboli:1999pt}
O.~J. Eboli, M.~Gonzalez-Garcia, S.~. Lietti, and S.~Novaes,
\newblock Phys.Lett. {\bf B478}, 199 (2000), arXiv:hep-ph/0001030.
%%CITATION = HEP-PH/0001030;%%

\bibitem{GonzalezGarcia:1999fq}
M.~Gonzalez-Garcia,
\newblock Int.J.Mod.Phys. {\bf A14}, 3121 (1999), arXiv:hep-ph/9902321.
%%CITATION = HEP-PH/9902321;%%

\bibitem{DeRujula:1991se}
A.~De~Rujula, M.~Gavela, P.~Hernandez, and E.~Masso,
\newblock Nucl.Phys. {\bf B384}, 3 (1992).
%%CITATION = NUPHA,B384,3;%%

\bibitem{Alam:1997nk}
S.~Alam, S.~Dawson, and R.~Szalapski,
\newblock Phys.Rev. {\bf D57}, 1577 (1998), arXiv:hep-ph/9706542.
%%CITATION = HEP-PH/9706542;%%

\bibitem{Espinosa:2012ir}
J.~Espinosa, C.~Grojean, M.~Muhlleitner, and M.~Trott,
\newblock JHEP {\bf 1205}, 097 (2012), arXiv:1202.3697.
%%CITATION = ARXIV:1202.3697;%%

\bibitem{Ellis:2012rx}
J.~Ellis and T.~You,
\newblock (2012), arXiv:1204.0464.
%%CITATION = ARXIV:1204.0464;%%

\bibitem{Carmi:2012yp}
D.~Carmi, A.~Falkowski, E.~Kuflik, and T.~Volansky,
\newblock (2012), arXiv:1202.3144.
%%CITATION = ARXIV:1202.3144;%%

\bibitem{Azatov:2012bz}
A.~Azatov, R.~Contino, and J.~Galloway,
\newblock JHEP {\bf 1204}, 127 (2012), arXiv:1202.3415.
%%CITATION = ARXIV:1202.3415;%%

\bibitem{Klute:2012pu}
M.~Klute, R.~Lafaye, T.~Plehn, M.~Rauch, and D.~Zerwas,
\newblock (2012), arXiv:1205.2699.
%%CITATION = ARXIV:1205.2699;%%

\bibitem{Bonnet:2011yx}
F.~Bonnet, M.~Gavela, T.~Ota, and W.~Winter,
\newblock Phys.Rev. {\bf D85}, 035016 (2012), arXiv:1105.5140.
%%CITATION = ARXIV:1105.5140;%%

\bibitem{Arzt:1994gp}
C.~Arzt, M.~Einhorn, and J.~Wudka,
\newblock Nucl.Phys. {\bf B433}, 41 (1995), arXiv:hep-ph/9405214.
%%CITATION = HEP-PH/9405214;%%

\bibitem{Aad:2012mr}
ATLAS Collaboration, G.~Aad {\em et~al.},
\newblock (2012), arXiv:1205.2531.
%%CITATION = ARXIV:1205.2531;%%

\bibitem{Martelli:2012ea}
CMS collaboration, A.~Martelli and f.~t.~C. collaboration,
\newblock (2012), arXiv:1201.4596.
%%CITATION = ARXIV:1201.4596;%%

\bibitem{Hagiwara:1986vm}
K.~Hagiwara, R.~Peccei, D.~Zeppenfeld, and K.~Hikasa,
\newblock Nucl.Phys. {\bf B282}, 253 (1987).
%%CITATION = NUPHA,B282,253;%%

\bibitem{Nakamura:2010zzi}
Particle Data Group, K.~Nakamura {\em et~al.},
\newblock J.Phys.G {\bf G37}, 075021 (2010).
%%CITATION = JPHGB,G37,075021;%%

\bibitem{Eboli:2010qd}
O.~Eboli, J.~Gonzalez-Fraile, and M.~Gonzalez-Garcia,
\newblock Phys.Lett. {\bf B692}, 20 (2010), arXiv:1006.3562.
%%CITATION = ARXIV:1006.3562;%%

\bibitem{Fogli:2002pt}
G.~Fogli, E.~Lisi, A.~Marrone, D.~Montanino, and A.~Palazzo,
\newblock Phys.Rev. {\bf D66}, 053010 (2002), arXiv:hep-ph/0206162.
%%CITATION = HEP-PH/0206162;%%

\bibitem{GonzalezGarcia:2007ib}
M.~Gonzalez-Garcia and M.~Maltoni,
\newblock Phys.Rept. {\bf 460}, 1 (2008), arXiv:0704.1800.
%%CITATION = ARXIV:0704.1800;%%

\bibitem{Dittmaier:2011ti}
LHC Higgs Cross Section Working Group, S.~Dittmaier {\em et~al.},
\newblock (2011), arXiv:1101.0593.
%%CITATION = ARXIV:1101.0593;%%

\bibitem{Espinosa:2012vu}
J.~R. Espinosa, M.~Muhlleitner, C.~Grojean, and M.~Trott,
\newblock (2012), arXiv:1205.6790.
%%CITATION = ARXIV:1205.6790;%%

\bibitem{Raidal:2011xk}
M.~Raidal and A.~Strumia,
\newblock Phys.Rev. {\bf D84}, 077701 (2011), arXiv:1108.4903.
%%CITATION = ARXIV:1108.4903;%%

\bibitem{Alwall:2011uj}
J.~Alwall, M.~Herquet, F.~Maltoni, O.~Mattelaer, and T.~Stelzer,
\newblock JHEP {\bf 1106}, 128 (2011), arXiv:1106.0522.
%%CITATION = ARXIV:1106.0522;%%

\bibitem{Christensen:2008py}
N.~D. Christensen and C.~Duhr,
\newblock Comput.Phys.Commun. {\bf 180}, 1614 (2009), arXiv:0806.4194.
%%CITATION = ARXIV:0806.4194;%%

\bibitem{Pukhov:1999gg}
A.~Pukhov {\em et~al.},
\newblock (1999), arXiv:hep-ph/9908288.
%%CITATION = HEP-PH/9908288;%%

\bibitem{Boos:2004kh}
CompHEP Collaboration, E.~Boos {\em et~al.},
\newblock Nucl.Instrum.Meth. {\bf A534}, 250 (2004), arXiv:hep-ph/0403113.
%%CITATION = HEP-PH/0403113;%%

\bibitem{Arnold:2011wj}
K.~Arnold {\em et~al.},
\newblock (2011), arXiv:1107.4038.
%%CITATION = ARXIV:1107.4038;%%

\end{thebibliography}

%%%%%%%%%%%%%%%%%%%%%%%%%%%%%%%%%%%%%%%%%%%%%%%%%%%%%%%%%%%%%%%%%%%%%%
\end{document}